\documentclass[twocolumn,showpacs,preprintnumbers,amsmath,amssymb]{revtex4}
\usepackage{amsmath,amssymb,color,epsfig,latexsym,natbib,graphicx,dcolumn}
\usepackage{bm}
\bibpunct{[}{]}{,}{n}{,}{,}             
  
\def\DEL#1{{\textcolor{green}{}}}         

\def\ni{\noindent}

\newcommand{\bb}{\textbf{b}}

\newcommand{\bB}{\textbf{B}}

\newcommand{\bk}{\textbf{k}}
\newcommand{\bp}{\textbf {p}}
\newcommand{\bq}{\textbf {q}}

\newcommand{\bv}{\textbf{v}}

\newcommand{\bx}{\textbf{x}}

\newcommand{\rem}[1]{}

\newcommand\vecp[1]{\vec{#1}}                   

\newcommand{\be}{\begin{equation}}
\newcommand{\ee}{\end{equation}}

\def\bB0{\vecp{B}_0}   

\begin{document}
\title{\bf Spectral Modeling of Magnetohydrodynamic Turbulent Flows} 
\author{J. Baerenzung}
\affiliation{TNT/NCAR, P.O. Box 3000, Boulder, Colorado 80307-3000, U.S.A.\\
Universit\'e de Nice-Sophia Antipolis, CNRS UMR 6202, Observatoire de la C\^ote d'Azur, B.P. 4229, 06304 Nice Cedex 4, France }
\author{H. Politano}
\affiliation{Universit\'e de Nice-Sophia Antipolis, CNRS UMR 6202, Observatoire de la C\^ote d'Azur, B.P. 4229, 06304 Nice Cedex 4, France }
\author{Y. Ponty}
\affiliation{Universit\'e de Nice-Sophia Antipolis, CNRS UMR 6202, Observatoire de la C\^ote d'Azur, B.P. 4229, 06304 Nice Cedex 4, France }
\author{A. Pouquet}
\affiliation{TNT/NCAR, P.O. Box 3000, Boulder, Colorado 80307-3000, U.S.A.}

\begin{abstract}
We present a dynamical spectral model for Large Eddy 
Simulation of the incompressible magnetohydrodynamic (MHD) equations based on the 
Eddy Damped Quasi Normal Markovian approximation. 
This model extends classical spectral Large Eddy Simulations for the Navier-Stokes equations to incorporate general (non Kolmogorovian) spectra as well as eddy noise. We derive the model for MHD and show that introducing a new eddy-damping time for the dynamics of spectral tensors in the absence of equipartition between the velocity and magnetic fields leads to better agreement with direct numerical simulations, an important point for dynamo computations. 
\end{abstract}
\maketitle

\section{Introduction}
Magnetic fields permeate the universe.
If kinetic effects such as Hall current, ambipolar drift of anisotropic pressure tensor, may be prevalent at small scales, the large-scales can be 
described in the magnetohydrodynamic (MHD) approximation. For example, electric fields and ionospheric currents play a dynamic role in the 
evolution of the atmosphere above 100 km, and the input of energy from
the magnetosphere during magnetic storms can affect the thermosphere and 
ionosphere on global scales.

MHD has
many similarities with Navier--Stokes (NS) turbulence: recall the Batchelor
analogy between vorticity and induction, both undergoing stretching through
velocity gradients (see equation (\ref{mhd_eq})). On that basis, one can 
conjecture that the energy spectrum will be of the Kolmogorov type, as 
in fact observed in numerical simulations of a decaying flow (\cite{grappin83} 
and references therein), as well as in the Solar Wind \cite{matt}. However, 
Iroshinkov and Kraichnan (IK hereafter) hypothesized that the slowing-down 
of nonlinear transfer by Alfv\'en waves would alter the energy spectrum 
\cite{IKI}, predicting a spectrum $\sim k^{-3/2}$, as recently observed in 
numerical simulations \cite{maron,mueller_grappin,1536WT} and in Solar Wind 
observations \cite{podesta}.
The Lagrangian renormalized approximation also gives spectra compatible with 
the IK model \cite{yoshida}. These spectra are isotropic, but the presence 
of a strong quasi-uniform magnetic field ${\bf{B}_0}$ at large scale renders 
the dynamic anisotropic. One can compute exactly the reduced dynamics in 
that case \cite{galtier00}, using weak turbulence theory; the emerging energy 
spectrum $\sim|{\bf{k}_{\perp}|^{-2}}$ , where ${\bf{k}_{\perp}}$ refers to 
wavevectors perpendicular to ${\bf{B}_0}$ (note that the isotropization of 
such a spectrum is compatible with the IK spectrum). Note also that a weak 
turbulence spectrum was observed in the magnetosphere of Jupiter \cite{saur}, 
the evidence stemming from an analysis of Galileo spacecraft data. A weak 
turbulence spectrum also obtained as well in a large numerical simulation of 
MHD turbulence in three dimensions \cite{1536WT} at a magnetic Taylor 
Reynolds number of $\sim 1700$ at scales smaller than where the isotropic
IK spectrum is observed.

Both the terrestrial and Jovian magnetospheric plasmas as well as
the solar wind, the solar atmosphere, and the interstellar medium 
are highly turbulent conducting 
compressible flows sustaining magneto-acoustic wave propagation and a 
better understanding of their dynamics, leading for example to star 
formation, requires adequate tools for modeling them. Furthermore, there is 
currently a surge of interest for achieving an experimental dynamo (the growth 
of a seed magnetic field through fluid motions, see \cite{special}). 
In the case of liquid metals, or the fluid core of the Earth 
\cite{lathrop,bourgoin02,petrelis03} or the solar convection zone, 
the magnetic Prandtl number  is very small ($10^{-5}$ or less); hence, 
the dynamo instability occurs in a turbulent flow and modeling the 
turbulence in order to study this phenomenon is in order 
\cite{ponty_04,ponty_05}.

There are few models for MHD (see e.g. the recent review in \cite{pps}), 
comparatively to the fluid case where the engineering community has been 
driving a vigorous research agenda. In that light, this paper aims at 
developing such a model, in the context of a spectral approach
following the work of Chollet and Lesieur \cite{chollet_lesieur}
for the fluid case, using two-point closure of turbulent flows. We give 
the basic equations in the next section and then move on in Section \ref{modeling} 
to recall the EDQNM closure formulation for MHD. New triad relaxation times
 are introduced in Section \ref{deriv} and first tested in Section \ref{test}.
The case of a random flow for two values of the magnetic Prandtl number 
$P_M$ are treated respectively in Section \ref{random} ($P_M=1$) and 
Section \ref{random2} ($P_M=0.1$), and the deterministic Orszag-Tang flow in 
three dimension is analyzed in Section \ref{OT}.
Finally, Section \ref{conclu} is the conclusion and  two technical appendices 
(\ref{appenA} and \ref{appenB}) are given at the end.

\section{Magnetohydrodynamic equations}\label{eqs}
Let us consider the Fourier transform of the velocity $\bv(\bx,t)$ and
the magnetic $\textbf{B}(\bx,t)$ at wavevector $\bk$:
\begin{eqnarray}
   \textbf{v}(\textbf{k},t) &=& \int_{-\infty}^\infty \textbf{v}(\textbf{x},t) 
e^{-i\textbf{k}.\textbf{x}} \textbf{dx} \\
   \textbf{B}(\textbf{k},t) &=& \int_{-\infty}^\infty \textbf{B}(\textbf{x},t) 
e^{-i\textbf{k}.\textbf{x}} \textbf{dx} .
\end{eqnarray}
The MHD equations describe the time evolution of a conducting fluid velocity 
field coupled to a magnetic field. They derive from Maxwell's equations with 
the assumption that velocities are sub-relativistic, hence the displacement 
current can be neglected \cite{parker_book,moffatt}. In terms of the Fourier 
coefficients of the velocity and the magnetic components, the MHD  equations 
with constant unit density read:
\begin{eqnarray}
\left(\frac{\partial}{\partial t}+\nu k^2 \right) \bv(\bk,t)
&=& \textbf{t}^V(\bk,t)\\
\left(\frac{\partial}{\partial t}+\eta k^2 \right) \bb(\bk,t)
&=& \textbf{t}^M(\bk,t) \ ,
\label{mhd_eq} \end{eqnarray}
\ni with ${\bk \cdot \bv} = {\bf 0}$ in the incompressible case and
 $\bk \cdot {\bf b}=0$
indicating the lack of magnetic monopoles in the classical approximation.
Here, ${\bf b} = {\bf B} /\sqrt {\mu_0 \rho_0}$ is the Alfv\'en velocity, with
$\mu_0$ the permeability, and $\rho_0$ the uniform density (taken equal to 
unity); $\eta$ is the 
magnetic diffusivity, $\nu$ the kinematic viscosity, and 
$\textbf{t}^V(\bk,t)$ and $\textbf{t}^M(\bk,t)$ 
are bilinear operators for energy transfer written as:
\begin{eqnarray}
{t_\alpha}^V(\bk,t)
&=&-iP_{\alpha\beta}(\bk)k_\gamma \sum_{\textbf{p}+ \textbf{q} = 
\textbf{k}} v_\beta(\bp,t) v_\gamma(\bq,t) \nonumber\\  
& & + iP_{\alpha\beta}(\bk)k_\gamma \sum_{\textbf{p}+ \textbf{q} = 
\textbf{k}} b_\beta(\bp,t)b_\gamma(\bq,t)\\
{t_\alpha}^M(\bk,t)
&=& -i\delta_{\alpha\beta}k_\gamma \sum_{\textbf{p}+ \textbf{q} = 
\textbf{k}} b_\beta(\bp,t)
v_\gamma(\bq,t) \nonumber \\
& &  -i\delta_{\alpha\beta}k_\gamma \sum_{\textbf{p}+ \textbf{q} = 
\textbf{k}} b_\beta(\bq,t)v_\gamma(\bp,t) \ ,
\label{mhd_eq} \end{eqnarray}
with $P_{\alpha\beta}(\bk)=\delta_{\alpha\beta} - {k_\alpha k_\beta}/{k^2}$
a projector that allows to take the pressure term of the velocity equation 
into account via the Poisson formulation. The magnetic Prandtl 
number is defined as $P_M=\nu / \eta$.
Finally note that, in the absence of dissipation ($\nu=0=\eta$), the total energy
$E_T=0.5<v^2+b^2>$, the correlation between the velocity and the magnetic field $H_C=<{\bf v}\cdot {\bf b}>$ and the magnetic helicity $<{\bf A} \cdot {\bf b}>$ (with ${\bf b}=\nabla \times {\bf A}$) are invariants of the ideal 
MHD equations in three dimensions.

\section{Spectral modeling}\label{modeling}
\subsection{The original EDQNM closure}
The Large Eddy Simulation model (LES) derived in \cite{LESPH} (Paper I hereafter) is now extended
to the MHD equations in its non-helical version (LES-P). As a first step, a spectral filtering of the
equations is realized; this operation consists in the truncation
of all veloctity and magnetic components at a wave-vector $\bk$ such
that $|\bk | > k_c$ where $k_c$ is a cut-off wavenumber.
In an intermediate zone lying between $k_c$ and $3k_c$ both kinetic and magnetic 
energy spectra are assumed to behave as power-laws followed by an exponential
decrease:
\begin{eqnarray}
E^V(k,t) & = & E_0^V k^{-\alpha_E^V} e^{-\delta_E^V k}, \quad  k_c\le k<3k_c \label{fit_Ev}\\
E^M(k,t) & = & E_0^M k^{-\alpha_E^M} e^{-\delta_E^M k}, \quad  k_c\le k<3k_c \ , \label{fit_Em} 
\end{eqnarray}
where $\alpha_E^V$, $\delta_E^V$, $E_0^V$, and $\alpha_E^M$, $\delta_E^M$, 
$E_0^M$, are evaluated at each time step of the numerical simulations,
through a mean square fit of 
the resolved kinetic and magnetic energy spectra. 
In a second step one can write the modeled MHD equations as:
\begin{eqnarray}
\left[\partial_t + \left(\nu\left(k|k_c,t\right) +\nu k^2\right) \right]
v_\alpha(\textbf{k},t) 
&=& t_\alpha^{V<}(\bk,t) \\
\left[\partial_t + \left(\eta\left(k|k_c,t\right) +\eta k^2\right) \right]
b_\alpha(\textbf{k},t) 
&=& t_\alpha^{M<}(\bk,t)
\label{mhdmodel}
\end{eqnarray}
where the $<$ symbol indicates that the nonlinear transfers are
integrated on the truncated domain
such as $\bp +\bq = \bk$ with $|\bp|, |\bq| < k_c$. 
The quantities $\nu\left(k|k_c,t\right)$
and $\eta\left(k|k_c,t\right)$ which are respectively called
eddy viscosity and magnetic eddy diffusivity are expressed as:
\begin{eqnarray}
\nu(k|k_c,t)\!\!\! &=& \!\!\! - \iint_{\Delta^>}\!\!\!\theta_{_{kpq}}
\frac{\left(S_2^V(k,p,q,t) + S_{4}^V(k,p,q,t) \right)}{2 k^2E^V(k,t)}
dpdq \nonumber \\
& & \\
\eta(k|k_c,t)\!\!\! &=& \!\!\! - \iint_{\Delta^>}\!\!\!\theta_{_{kpq}}
\frac{\left(S_2^M(k,p,q,t) + S_{4}^M(k,p,q,t) \right)}{2 k^2E^M(k,t)}
dpdq \nonumber \\
\end{eqnarray}
(see Paper I for more details). Here the $S_i^{V,M}(k,p,q,t)$ terms (see Appendix A), correspond to the
absorption terms of the Eddy Damped Quasi Normal Markovian (EDQNM) 
nonlinear transfer, leading in particular to turbulent eddy diffusivities
(see e.g. \cite{KN} for the MHD case). 
$\Delta^>$ is the integration domain on $k$, $p$, $q$, triangles such 
as $p$ and or $q$ are larger than $k_c$ and both $p$ and $q$ are 
smaller than $3k_c$.

Finally, to take into account the effect of the emission (eddy-noise) terms of the EDQNM nonlinear 
transfer (i.e. $S_1^V(k,p,q,t)$, $S_3^V(k,p,q,t)$, $S_1^M(k,p,q,t)$, and
$S_3^M(k,p,q,t)$), we use a reconstruction field procedure
which also enables to partialy rebuild the phase 
relationships between the three spectral 
components of each velocity and magnetic fields, as explained in Paper I \cite{LESPH}.


\subsection{First numerical tests} \label{first}

We first implemented our LES model with the EDQNM equations
of Pouquet et al \cite{PFL} (see also \cite{KN}). 
In this formulation the triad-relaxation time
$\Theta_{kpq}$ (see Appendix A) takes three characteristic times
into account:

\noindent - a (combined) dissipation time $\tau_D$ defined as:
\begin{equation}
\tau_D^{-1}(k) = \left(\nu+\eta \right) k^2 \ ,
\label{td_pfl} \end{equation}

\noindent - a nonlinear time $\tau_{S}$ expressed as:
\begin{equation}
\tau_S^{-1}(k) =  \lambda\left[\int_0^k q^2 \left[E^V(q)+E^M(q)\right]dq 
\right]^{\frac{1}{2}} \ ,
\label{tnl_pfl} \end{equation}

\noindent - and an Alfv\'en time $\tau_{A}$ which reads:
\begin{equation}
\tau_A^{-1}(k) =  \left(\frac{2}{3} \right)^{\frac{1}{2}}
k\left[\int_0^k E^M(q)dq \right]^{\frac{1}{2}} \ .
\label{ta_pfl} \end{equation}

\begin{figure}[ht]
\includegraphics[width=7cm, height=50mm]{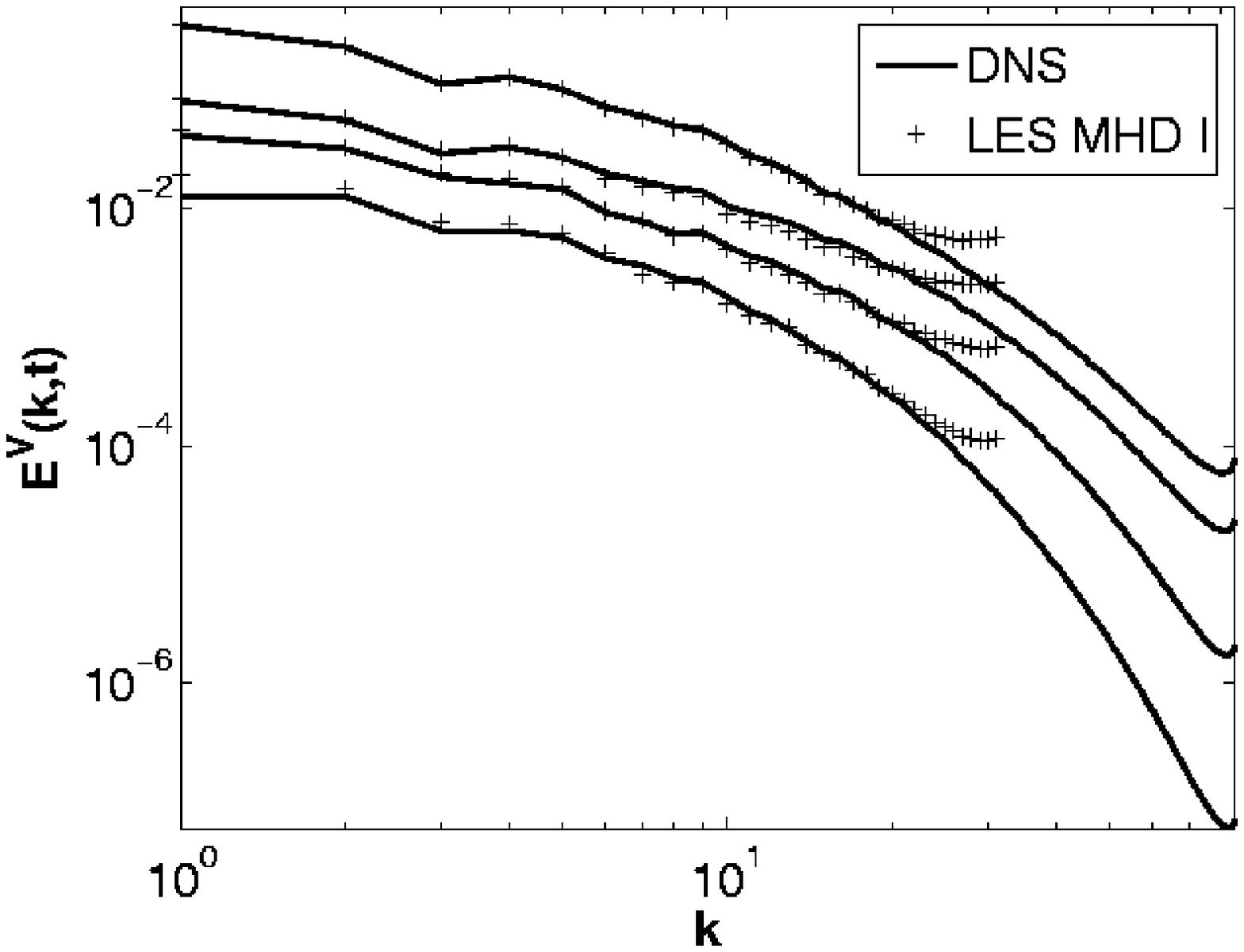}
\includegraphics[width=7cm, height=50mm]{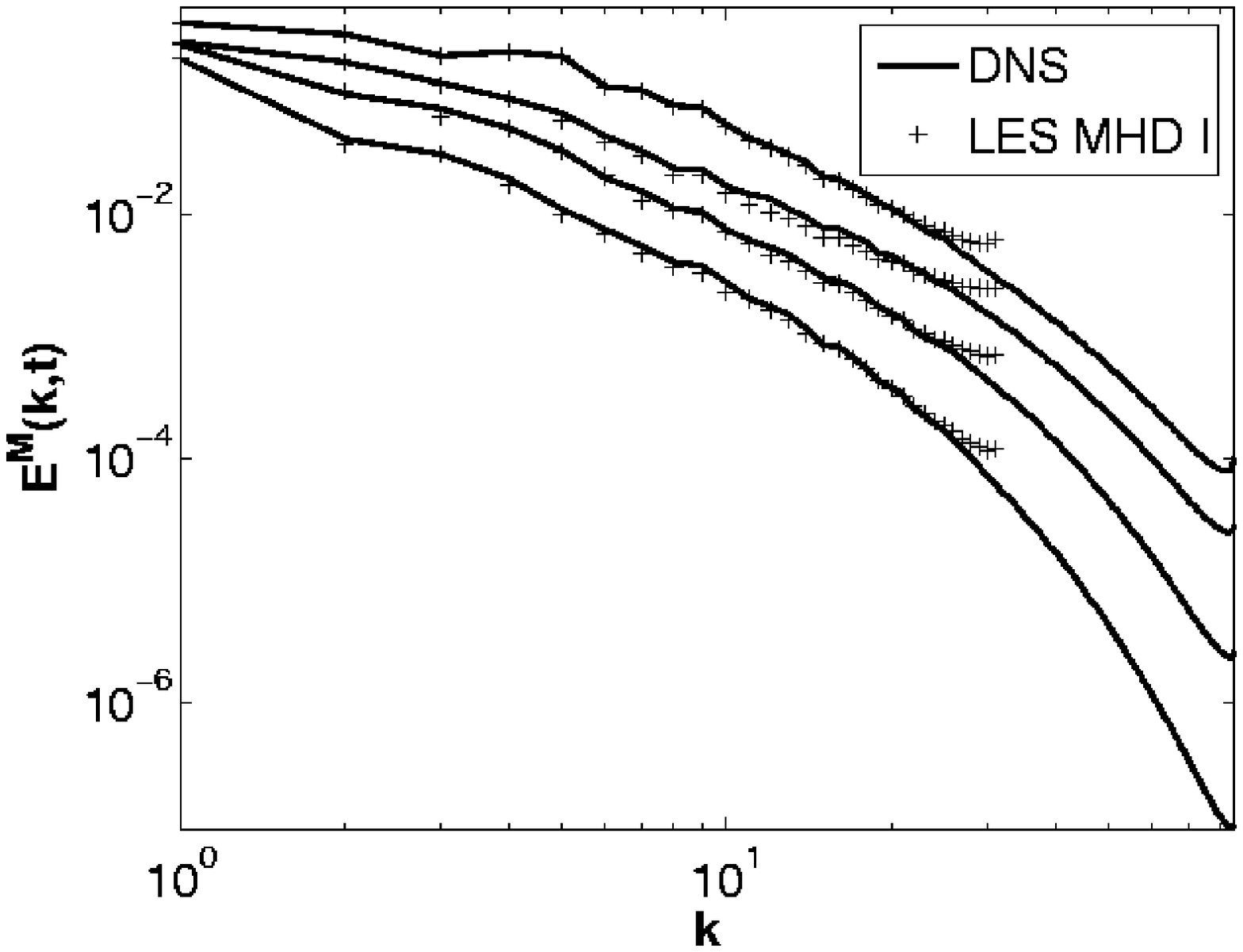}
\caption{Kinetic (top) and magnetic (bottom) energy spectra
at time $t=1$, $t=3$, $t=5$, and $t=10$ from upper 
to lower curves for data  
{\bf I} ($256^3$ DNS, solid line), and data {\bf II} 
($64^3$ LES MHD I, plus symbols).}
\label{compa_evo_pfl} 
\end{figure}

This constitutes a straightforward generalization of the EDQNM closure to the 
case of MHD flows (see for example \cite{leshouches}) in which two new times, 
specifically the Alfv\'en time and the diffusion time built on magnetic 
resistivity, are incorporated in a phenomenological manner.
A comparison of a simulation using this LES model (LES MHD I, or
run {\bf II} in Table~\ref{table1}), 
with a DNS simulation (run {\bf I} in Table~\ref{table1}) is shown in 
Fig.~\ref{compa_evo_pfl} (see next section for more information on the 
numerical procedure).  One sees that
both the kinetic and magnetic energy spectra are overestimated by the model
at scales close to the cut-off, indicative of an inadequate energy transfer 
in the model at these scales.
When evaluating numerically the different eddy dampings using eqns. 
(\ref{td_pfl}-\ref{ta_pfl}), one observes that the Alfv\'en time is 
almost one order of magnitude shorter than all other times, including 
the diffusion time at the smallest resolved scales (see \cite{JB_PHD}); 
this leads to an insufficient damping at the scales close to the cut-off. 
This is in part due to the dominance of the magnetic energy at large-scale; 
in that sense, it could be linked to the particular flow under study and  
parametric analyses of several flows will have to be performed in the future 
in order to fine-tune this MHD model. However, the discrepancy displayed 
in Fig.~\ref{compa_evo_pfl} could also be linked to the particular 
expression of eddy damping chosen in \cite{PFL}.
We are thus led to examine more closely the dynamics of energy transfer 
within the EDQNM framework.

\section{New relaxation times for EDQNM}\label{deriv}

We now analyze the precise structure of the equations leading to the EDQNM 
closure; this is done in Appendix \ref{appenB}. 
Note that a similar but more complex and systematic approach can be found in 
\cite{yoshida} in the context of the Lagrangian renormalization approximation. 
Our analysis results in the expression of three new eddy diffusivity times, 
with which one can build four different eddy-damping rates $\mu_{kpq}$, namely:
\begin{eqnarray}
\!\!\!\!\!\!\!\!\!\mu_{kpq}^{{VV}} \!\!\! &=& \!\!\! \tau_D^{{VV}^{-1}}(k,p,q) 
+ \tau_{NL}^{-1}(k,p,q) \ ,  \nonumber \\ 
\!\!\!\!\!\!\!\!\!\!\mu_{kpq}^{{VM}} \!\!\! &=& \!\!\! 
\tau_D^{{VM}^{-1}}(k,p,q) 
+ \tau_{NL}^{-1}(k,p,q)
+ \widetilde{\tau}_A^{-1}(k,p,q) \ ,  \nonumber \\
\!\!\!\!\!\!\!\!\!\!\mu_{kpq}^{{MV}} \!\!\! &=& \!\!\! 
\tau_D^{{MV}^{-1}}(k,p,q) 
+ \tau_{NL}^{-1}(k,p,q)
+ \widetilde{\tau}_A^{-1}(k,p,q)\ , \\ 
\!\!\!\!\!\!\!\!\!\!\mu_{kpq}^{{MM}} \!\!\! &=& \!\!\! 
\tau_D^{{MM}^{-1}}(k,p,q) 
+ \tau_{NL}^{-1}(k,p,q)
+ \widetilde{\tau}_A^{-1}(k,p,q)\, \label{new_td2}
\end{eqnarray}
where:
\begin{eqnarray}
\tau_{NL}^{-1}(k,p,q)&=& \tau_{NL}^{-1}(k) + \tau_{NL}^{-1}(p) 
+ \tau_{NL}^{-1}(q) \ , \\
\widetilde{\tau}_A^{-1}(k,p,q) &=& \widetilde{\tau}_A^{-1}(k) 
+ \widetilde{\tau}_A^{-1}(p) + \widetilde{\tau}_A^{-1}(q)\ .
\end{eqnarray}
with $\widetilde{\tau}_A^{-1}(k,p,q)$ based on a new Alfv\'en-like time, namely 
\begin{equation}
\widetilde{\tau}_A^{-1}(k) =  C_A \left(\frac{\int_0^k E^M(q)dq}
{\int_0^k E^V(q)dq}\right)^{\frac{1}{2}} 
\left[\int_0^k q^2 E^M(q)dq 
\right]^{\frac{1}{2}}. 
\end{equation}

Finally, from these eddy-damping rates we derive the corresponding 
triad-interaction times (as defined in Appendix A). For the EDQNM kinetic
energy equation, we obtain two different $\Theta$'s,
$\Theta_{kpq}^{VV}$ applied to $S_1^V(k,p,q,t)$
and $S_2^V(k,p,q,t)$, and  $\Theta_{kpq}^{MM}$ applied to $S_3^V(k,p,q,t)$
and $S_4^V(k,p,q,t)$. 

For the EDQNM magnetic energy equation we also obtain  
two triad-interaction time, namely  $\Theta_{kpq}^{MM}$
applied to $S_1^M(k,p,q,t)$
and $S_2^M(k,p,q,t)$ and $\Theta_{kpq}^{MV}$ applied to $S_3^M(k,p,q,t)$
and $S_4^M(k,p,q,t)$. 
Note that this formulation leads to
distinguish between the Joule and the viscous dissipation 
more systematically than in \cite{PFL},
and therefore to possibly better simulate flows at magnetic Prandtl numbers 
different from unity. Another difference is the special treatment
of the Lorentz force in the velocity equation.
We call LES MHD II the model that takes these new 
triad-relaxation times into account.

\section{Numerical set-up}\label{test}

In order to assess the ability of the model to reproduce the physics involved 
in MHD flows, we performed Direct Numerical Simulations (DNS) of 
the three-dimensional MHD equations, at a resolution of $256^3$ grid points, together with
computations based on our LES MHD formulation, but now using $64^3$ grid points.
We performed this comparative study from three different
simulations of freely decaying MHD flows. 
To test our model in a simple configuration, we first 
run a simulation at $P_M=1$ 
with random initial conditions and no correlation between
the velocity and the magnetic field (run {\bf I} for
the DNS, run {\bf II} for the LES MHD I, and run {\bf III} for the LES MHD II,
in Table~\ref{table1}).
Since the new eddy-damping times we derived allow for a clear distinction
between the kinematic viscosity and the magnetic dissipation,
we then simulated a flow with identical initial conditions 
but  with now $P_M=0.1$ (run {\bf IV} for the DNS, and run {\bf V} 
for the LES MHD II in Table~\ref{table1}). 
We recall that, in the work of \cite{PFL}, the EDQNM closure has been 
derived for the case when the cross-correlation (or cross-helicity)
spectrum $ H^C(k)=<\bv(\bk) \cdot \bb(\bk)(k)>$ is assumed to be identically 
zero in the presence of helicity (see \cite{grappin82,grappin83} for the non
helical case in the presence of velocity-magnetic field correlation). However,
for many flows, this quantity is non negligible; furthermore, it can be strong locally (in particular in the vicinity of vorticity and current sheets) even when the global correlation is close to zero \cite{jcp,matt07}. We thus performed as well
a simulation at $P_M=1$ for which the velocity 
and the magnetic field are significantly correlated, in order 
to see how our model may adapt to such a situation. 
We chose the so-called three-dimensional Orszag-Tang flow (run {\bf VI} for the DNS, 
and run {\bf VII} for the LES MHD II in Table~\ref{table1}) for which
initially $2H^C/E^T=0.5$.

\noindent From all these simulations, we studied global flow quantities 
such as the total, kinetic and magnetic energies, as well as helicities, 
and the cross-correlation energy. We also analyzed the
spectral behaviors of these quantities.
\begin{table}
\caption{\label{table1}Parameters of the simulations. 
Initial conditions I.C., grid resolution $N$, kinematic viscosity 
$\nu$, and magnetic Prandtl number $P_M=\nu/\eta$, with $\eta$ the
magnetic diffusivity.}
\begin{ruledtabular}
\begin{tabular}{cccccc}
         &        & I.C.    &   $N$  & $\nu$      & $P_M$       \\
{\bf I}  & DNS    & Random  &  $256$ & $2.e^{-3}$ & $1$   \\
{\bf II} & LES I  & Random  &  $64$  & $2.e^{-3}$ & $1$   \\
{\bf III}& LES II & Random  &  $64$  & $2.e^{-3}$ & $1$   \\
{\bf IV} & DNS    & Random  &  $256$ & $8.e^{-4}$ & $0.1$   \\
{\bf V}  & LES II & Random  &  $64$  & $8.e^{-4}$ & $0.1$   \\
{\bf VI} & LES I  & Random  &  $64$  & $8.e^{-4}$ & $0.1$   \\
{\bf VII}& DNS    & OT      &  $256$ & $2.e^{-3}$ & $1$   \\
{\bf VIII}& LES II& OT      &  $64$  & $2.e^{-3}$ & $1$   \\
\end{tabular}
\end{ruledtabular}
\end{table}
\section{Random flow at ${\bf P_M=1}$} \label{random}

We first investigate the model behavior for a flow with random initial 
conditions, presenting no cross-correlation, and at magnetic Prandtl number of unity.
\subsection{Inter-comparison of models}
In this section, we compare the efficiency between the 
model that involves the eddy damping times stemming from \cite{PFL} (LES MHD I), and the model
where the new eddy-damping times derived in Appendix B are included 
(LES MHD II).
In Figure~\ref{diff_e_pfl_new}, we plot the relative 
difference of the kinetic and magnetic energy spectra computed from both 
LES models with the ones computed from the DNS. 
The spectra are chosen at time $t=1$, close to the time of maximum dissipation. 

\begin{figure}[ht]
\includegraphics[width=7cm, height=50mm]{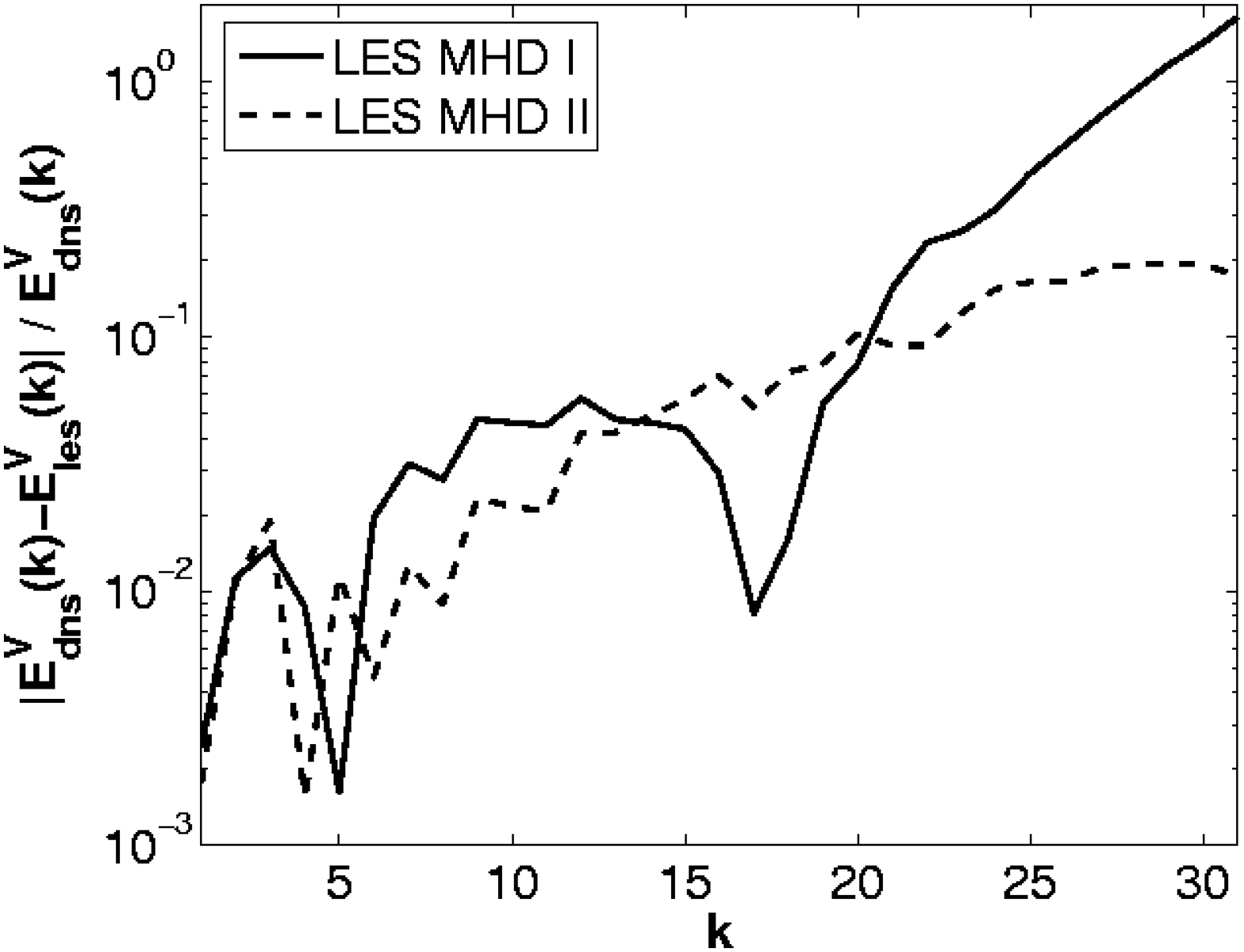}\\
\includegraphics[width=7cm, height=50mm]{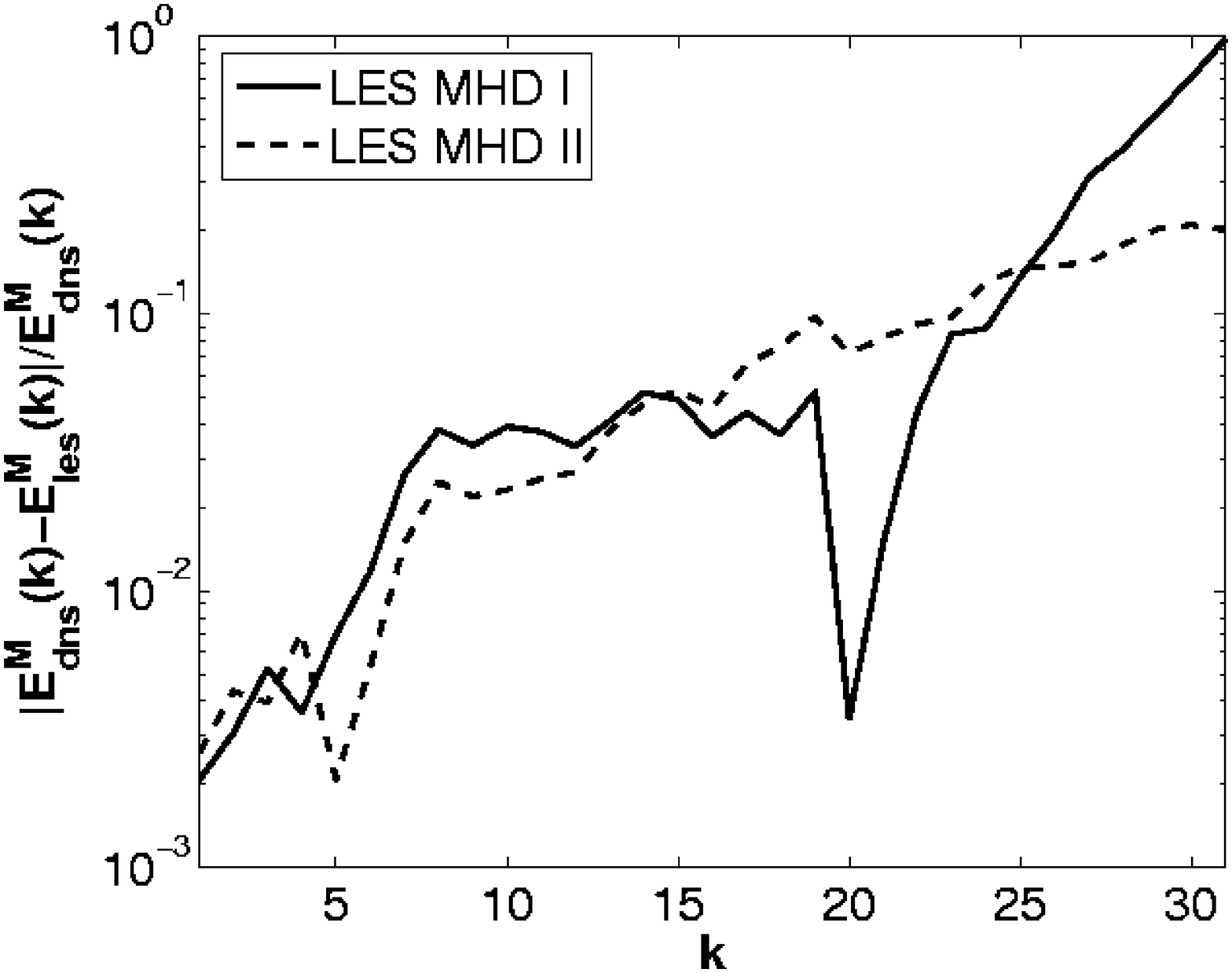}\\
\caption{Lin-log plots of the relative difference with DNS energy spectra for the velocity (top) and the
magnetic field (bottom) at time $t=1$, for runs  
{\bf II}  and {\bf III}, compared to the DNS, run I. Note the large error in 
LES I at large $k$.}
\label{diff_e_pfl_new} \end{figure}
   
At large scales (bewteen $k=0$ and $k\simeq 15$)
the LES MHD II model globally gives a better approximation of 
the kinetic and magnetic
energy spectra. Bewteen $k=14$ and $k=20$ for the kinetic energy spectra,
and between $k=15$ and $k=25$ for the magnetic energy spectra, the 
LES MHD I seems to give better results. This is due to the fact
that LES MHD I and  DNS spectra cross at
a wavenumber located inside these ranges. Finally, at small
scales, the LES MHD II data lead to a much better approximation
than the LES MHD I data. 
At different times, the comparison between LES MHD I and 
LES MHD II results leads to similar results (not shown). We therefore focus
our study on the LES MHD II model for the remainder of the paper.
\subsection{Global quantities}
\begin{figure}[ht]
\includegraphics[width=7cm, height=50mm]{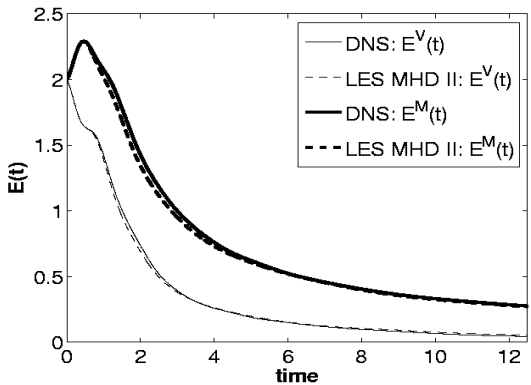}
\caption{Temporal evolution of the kinetic and magnetic energy for runs  
{\bf I} ($256^3$ DNS), and {\bf III} ($64^3$ LES MHD II). }
\label{compa_evt_ebt_pr1} \end{figure}

We here study the time evolution of the global kinetic, $E^V(t)$, and
magnetic, $E^M(t)$, energies
for runs {\bf I} and {\bf III}, as shown in Figure~\ref{compa_evt_ebt_pr1}).

One can observe that the modeled kinetic and magnetic energies 
both closely follow the DNS evolutions, although at short
times (between $t=1$ and $t=5$ for $E^M(t)$ and between $t=1$ and $t=3$ for
$E^V(t)$), the model slightly under-estimates them.

\begin{figure}[ht]
\includegraphics[width=7cm, height=50mm]{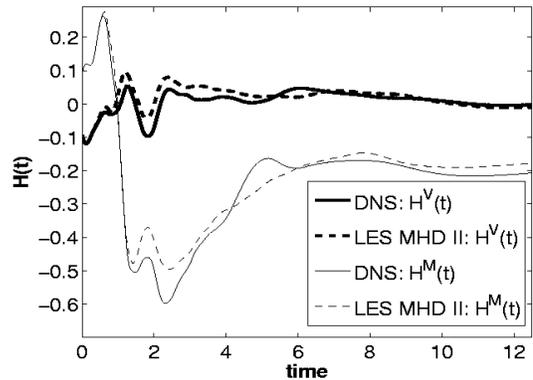}
\caption{Time evolution of the kinetic and magnetic helicities for runs  
{\bf I} ($256^3$ DNS), and {\bf III} ($64^3$ LES MHD II).}
\label{compa_hvt_hbt_pr1} \end{figure}

\noindent Since our field-reconstruction procedure uses the flow (kinetic and magnetic) helicities 
(even though the model itself does not take into account at this stage the helical contributions 
to evaluate the transport coefficient), we plot in Fig.~\ref{compa_hvt_hbt_pr1}
the time evolution of both kinetic and magnetic helicities (respectively
$H^V(t)$ and $H^M(t)$). 

One can notice that, even though both modeled kinetic and magnetic helicities 
do not exactly match the DNS results at each time, they remain close and 
reproduce the main DNS time fluctuations.
Note that the LES MHD I model provides similar results.

\noindent 
We do not present here the temporal evolution of 
the cross-helicity $H^C(t)$, since it is negligible when compared 
to the total magnetic and kinetic energy. Indeed, this correlation,
initially equal to zero, reaches a maximum value 
of $0.081$ for the DNS run,  and of $0.069$ for the LES MHD II run, to 
respectvely finish at a value of $0.051$ and $0.056$.

We now investigate the spectral behavior of our 
LES model by comparing the DNS and LES MHD II kinetic and
magnetic energy spectra at various dynamical times.

\begin{figure}[ht]
\includegraphics[width=7cm, height=50mm]{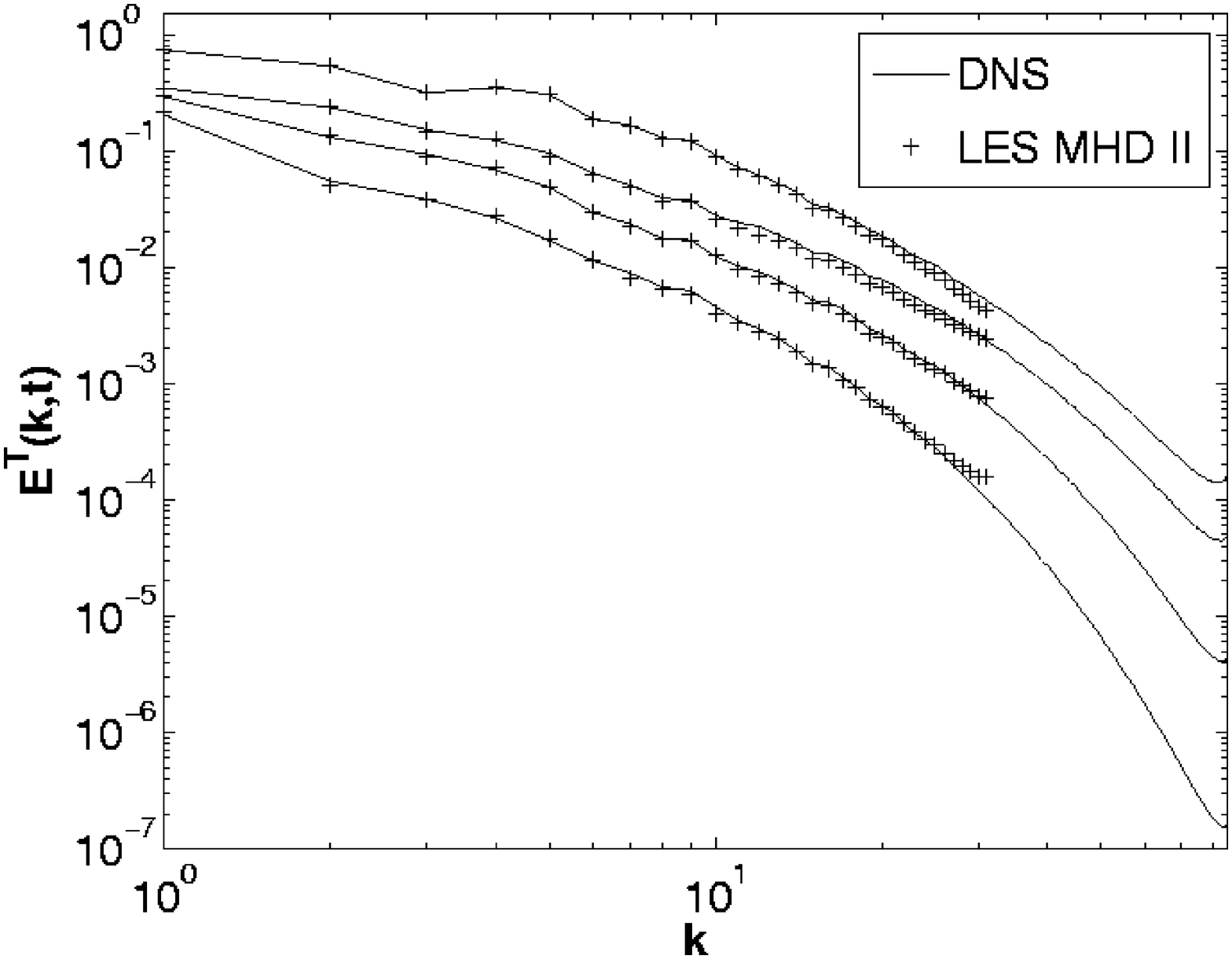}
\caption{Total energy spectra $E^T(k)=E^M(k)+E^V(k)$, at time $t=1$, $t=3$, 
$t=5$, and $t=10$ from top to bottom, for runs  
{\bf I} ($256^3$ DNS solid line), and {\bf III} ($64^3$ LES MHD II +).}
\label{compa_evo_et_pr1} \end{figure}

Figure~\ref{compa_evo_et_pr1} shows the total (kinetic plus magnetic) energy spectra 
$E^T(k)=E^M(k)+E^V(k)$ at
times  $t=1$, $t=3$, $t=5$, and $t=10$ obtained from DNS and  
LES MHD II computations.
At any wavenumber and  at any time, our LES MHD II model reproduces more correctly 
the DNS spectra than the LES MHD I does (see Fig.~\ref{compa_evo_pfl}).
It is clear that the spectral over-estimations 
at small scales obtained with this latter model is cured by 
the new formulation of the eddy-damping rates.
\section{Random flow at $\bf {P_M=0.1}$} \label{random2}
\begin{figure}[ht]
\includegraphics[width=7cm, height=50mm]{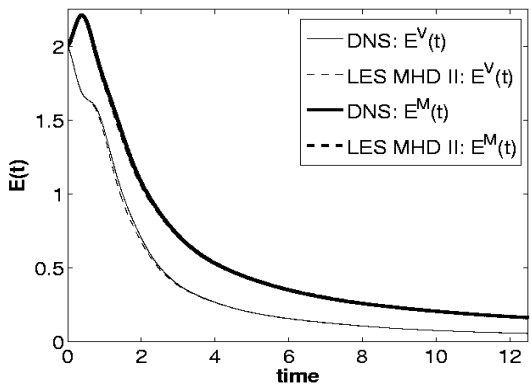}
\caption{Total kinetic and magnetic energy temporal evolution, for runs  
{\bf IV} ($256^3$ DNS), and {\bf V} ($64^3$ LES MHD II) at a magnetic
Prandtl number of $0.1$}
\label{compa_evt_ebt_pr01} \end{figure}

Since the new eddy-damping times involved in our LES MHD II model
allow for a more refined differenciation between the magnetic diffusivity
and the kinematic viscosity, we simulated a flow at 
a magnetic Prandtl number less than unity, namely $P_M=0.1$. 
In order to highlight the efficiency of the new damping times
to reproduce the flow dynamics, we compared 
both the LES MHD I and II data against the DNS results. 
For these simulations we kept identical flow initial
conditions as in the previous section.
A first comparison between the time evolution of the kinetic and magnetic
energies computed from a DNS, and a simulation
using the LES MHD II model, is plotted in Figure~\ref{compa_evt_ebt_pr01}.

One can observe that the model almost reproduces the exact
temporal evolution of both kinetic and magnetic energy. The evolution
of the kinetic and magnetic helicities (not shown) is also well-reproduced by the model. 
Once again, the cross-correlation remains weak all along the simulations; 
initially equal to zero, it reaches a maximum  value
of $0.056$ for the DNS, $0.057$ for the LES MHD I, and $0.057$ 
for the LES MHD II 
runs, before respective final values of $0.044$ (DNS), $0.046$ (LES MHD I), and $0.045$ 
(LES MHD II).

\begin{figure}[ht]
\includegraphics[width=7cm, height=50mm]{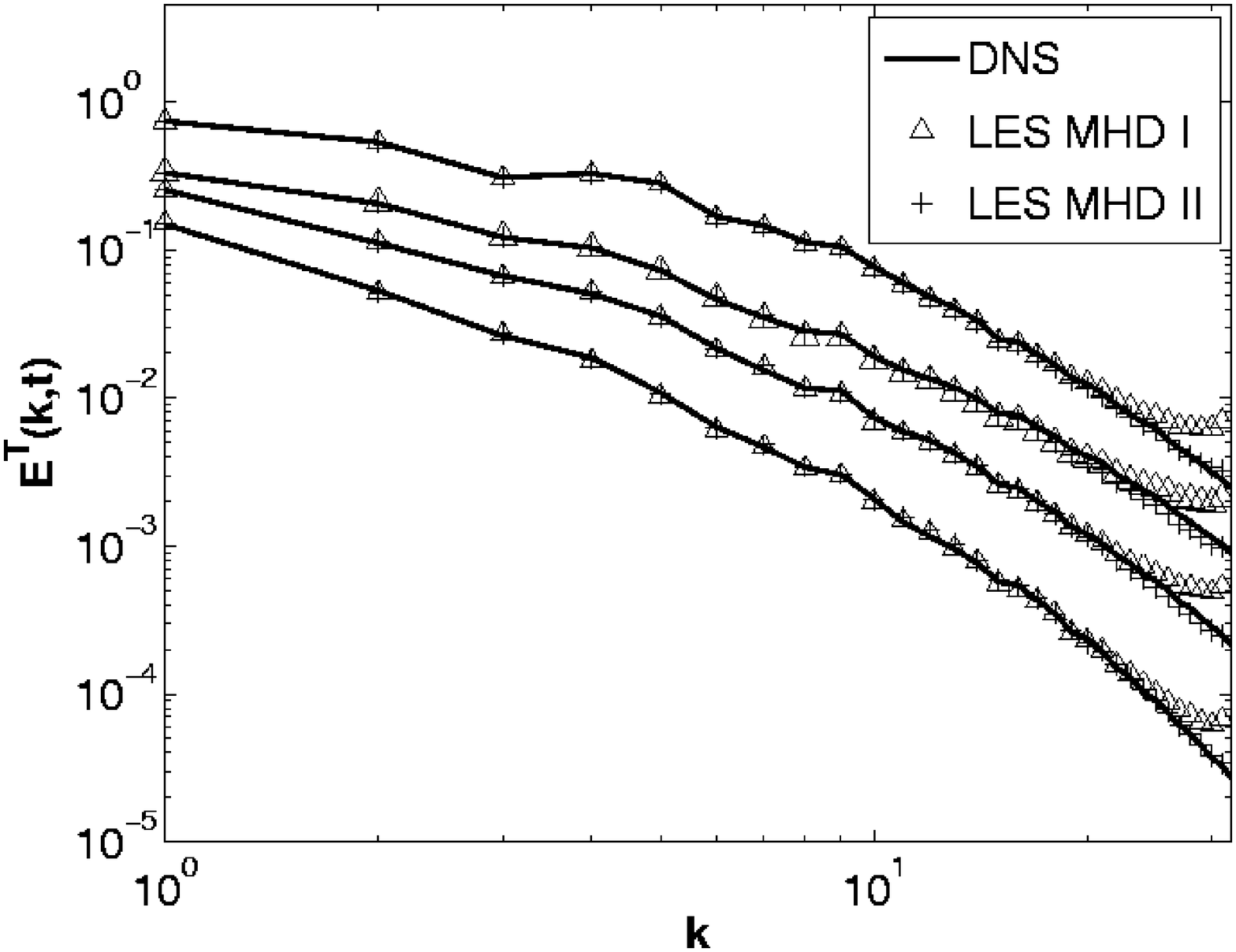}
\caption{Total energy spectra, at times $t=1$, $t=3$, 
$t=5$, and $t=10$ from top to bottom, for runs  
{\bf IV} ($256^3$ DNS, solid line), {\bf V} ($64^3$ LES MHD II, plusses), 
and {\bf VI} ($64^3$ LES MHD I, triangles) at $P_M=0.1$.}
\label{compa_evo_et_pr01} 
\end{figure}

We now present in Figure.~\ref{compa_evo_et_pr01} the total (kinetic plus magnetic) energy spectra
evolution at times $t=1$, $t=3$, $t=5$, and $t=10$,
obtained from DNS, LES MHD I, and LES MHD II data.
Although at small wavenumbers both LES models correctly reproduce the DNS spectra,
at large wavenumbers, strong differences appear 
among these various spectra.
Indeed, the LES MHD II results slightly
underestimate this range of the DNS spectra, whereas the
LES MHD I highly overestimates it. 
  
\section{Deterministic Orszag-Tang flow at ${\bf P_M=1}$} \label{OT}

For a majority of flows, the correlation between the velocity and the magnetic field (or cross-helicity)
is non negligible, leading to a slowed-down dynamics and energy spectra depending on the amount of correlation in the flow \cite{slow}.
It has also been observed that local patches of either aligned or anti-aligned velocity-magnetic field configurations
can be found both in the solar wind and in numerical simulations 
\cite{jcp,matt07}. We therefore decided to evaluate the ability
of our model to treat a flow with strong cross-correlation by examining 
the evolution
 of the so-called three-dimensional Orszag-Tang flow with
an initial global correlation $H^C(t=0)=1.63$ (to be
compared with the total kinetic and magnetic energy  $E^V(t=0)
=E^M(t=0)=2$). 

\subsection{Global quantities}

\begin{figure}[ht]
\includegraphics[width=7cm, height=50mm]{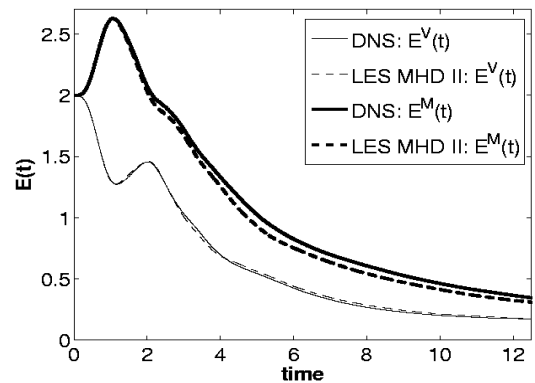}
\caption{Kinetic and magnetic energy evolution, for runs  
{\bf VII} ($256^3$ DNS solid line), and {\bf VIII} 
($64^3$ LES MHD II dashed line) with non-zero velocity-magnetic field 
correlation.}
\label{compa_evt_ebt_ot} \end{figure}

The kinetic energy evaluated with the LES MHD II 
fits with great accuracy to the kinetic energy obtained with the DNS
(see Fig.~\ref{compa_evt_ebt_ot}); however, the magnetic energy 
which is well-reproduced until $t=2$ departs measurably from the DNS data after
this time. 
\begin{figure}[ht]
\includegraphics[width=7cm, height=50mm]{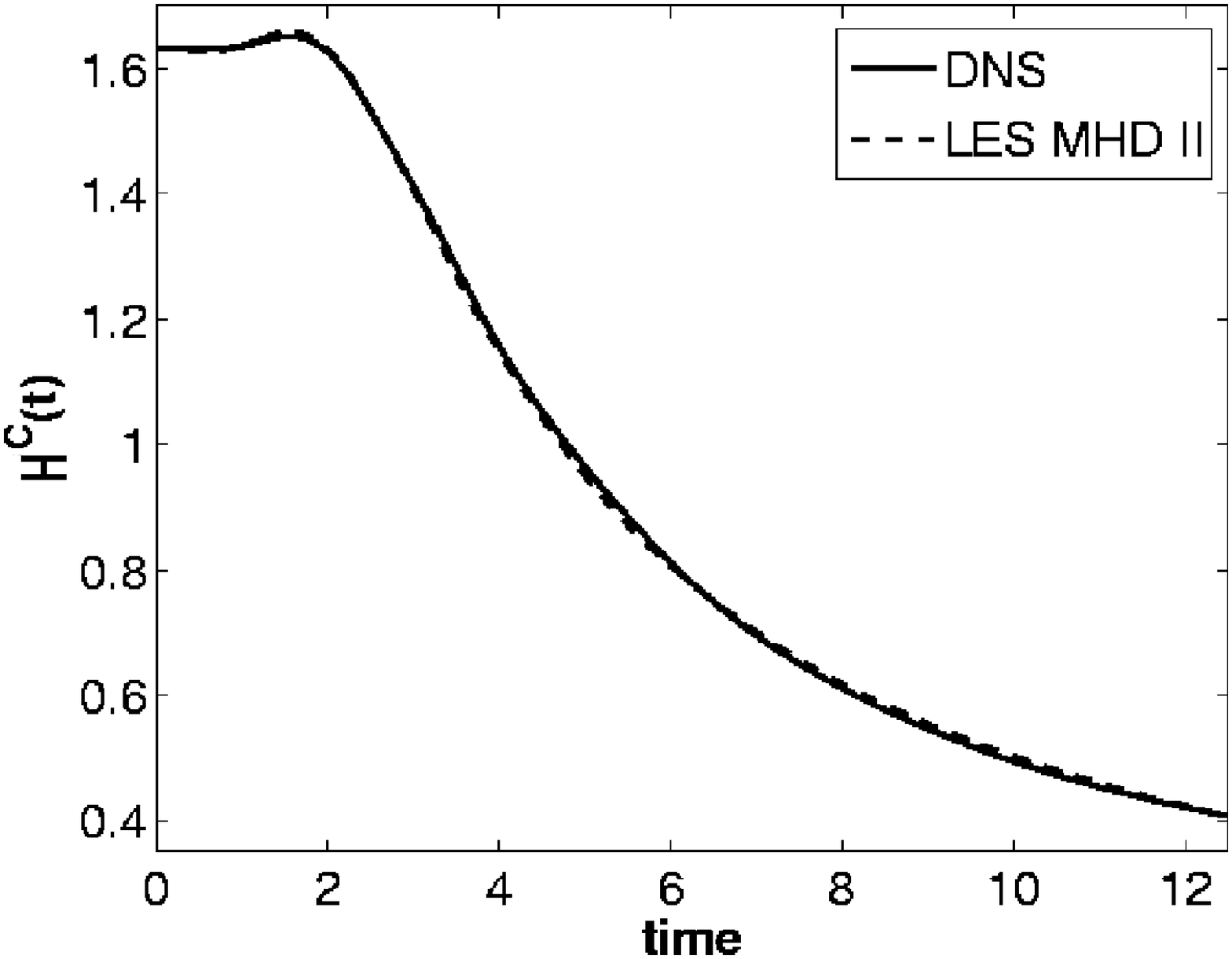}
\caption{Global velocity-magnetic field cross correlation, for runs  
{\bf VII} ($256^3$ DNS straight line), and {\bf VIII} 
($64^3$ LES MHD II dashed line).}
\label{compa_ect_ot} \end{figure}

\begin{figure}[ht]
\includegraphics[width=7cm, height=50mm]{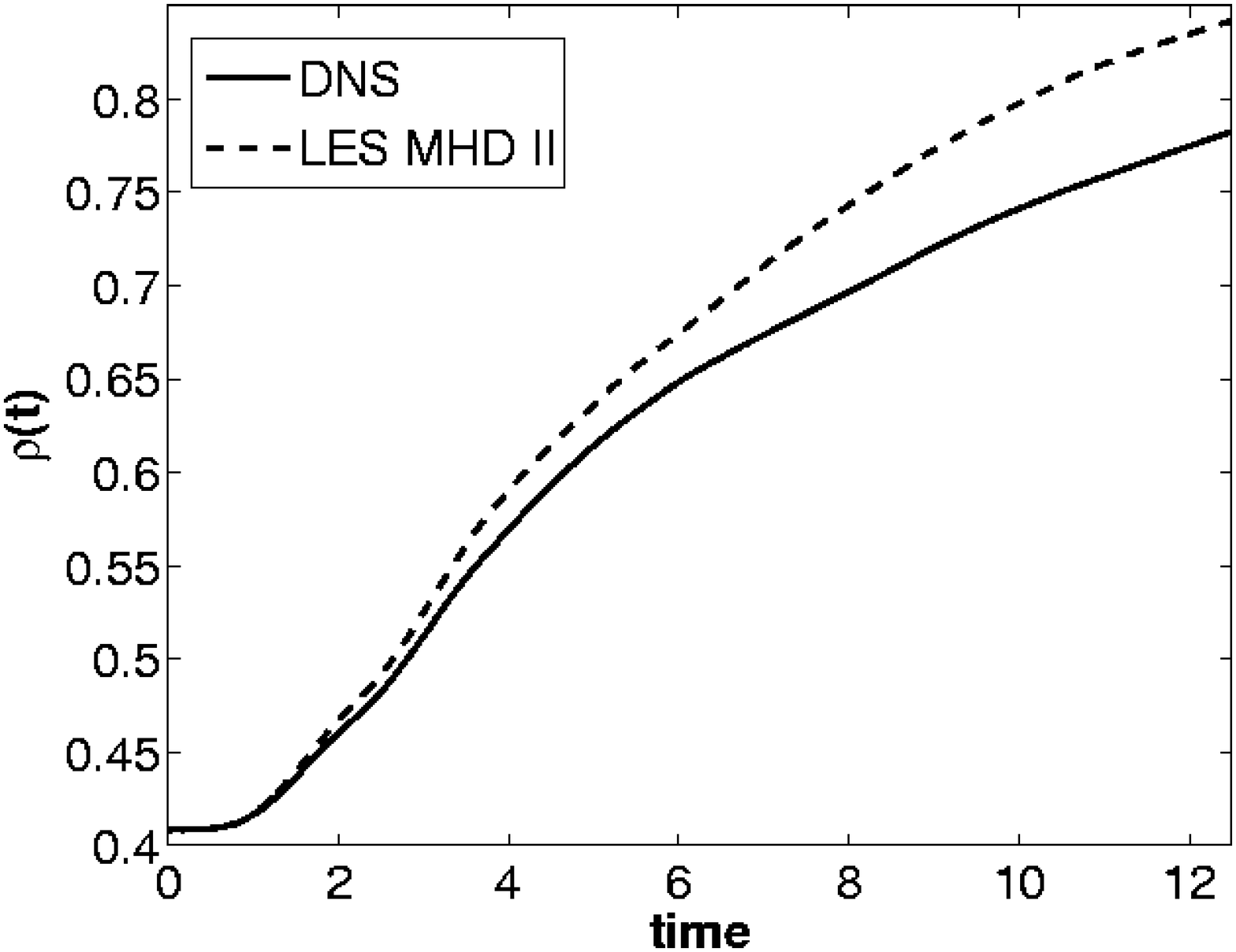}
\caption{Correlation coefficient $\rho(t)$, for runs  
{\bf VII} ($256^3$ DNS solid line), and {\bf VIII} 
($64^3$ LES MHD II dashed line).}
\label{compa_rho_ot} \end{figure}

\noindent 
The global cross-correlation, computed from either
DNS or LES MHD II data,  are quite close (see Fig.~\ref{compa_ect_ot}), 
demonstrating that although the model does not explicitly take this quantity into 
account, it still maintains a reliable evolution for it. However, the well-known 
temporal growth of the normalized cross-correlation coefficient 
$\rho(t)=H^C(t)/(E^V(t)+E^M(t))$ shown in  Fig.~\ref{compa_rho_ot} is not 
represented as accurately as either $E_T$ or $H^C$. This could be tentatively 
attributed to the fact that turbulent transport coefficients based on the 
velocity-magnetic field correlation itself would emerge from a complete model 
(as derived in \cite{grappin83}, see also \cite{grappin82}) 
the effect of which might be to dampen the correlation 
growth over time. Note that this discrepancy likely emerges 
from the less accurate representation of the magnetic energy itself, as 
displayed in Fig.~\ref{compa_evt_ebt_ot}.

\subsection{Spectral features} 

\begin{figure}[ht]
\includegraphics[width=7cm, height=50mm]{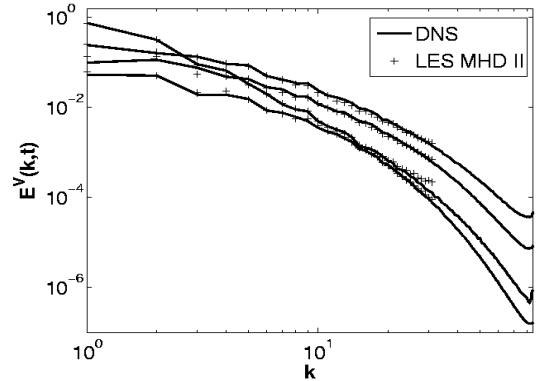}
\caption{Kinetic energy spectra, at times $t=1$, $t=3$, 
$t=5$, and $t=10$ from up to down, for data  
{\bf VII} ($256^3$ DNS straight line), and {\bf VIII} 
($64^3$ LES MHD II plusses).}
\label{compa_evo_ev_ot} \end{figure}

\begin{figure}[ht]
\includegraphics[width=7cm, height=50mm]{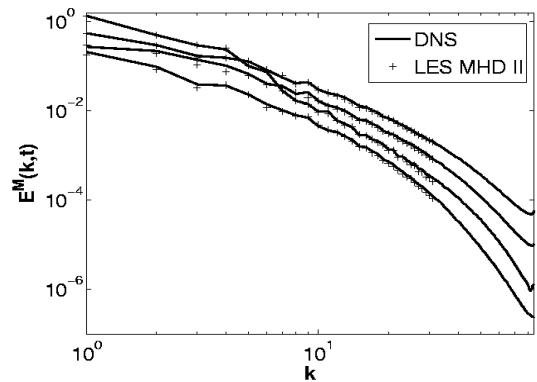}
\caption{Same as Fig. \ref{compa_evo_ev_ot} for magnetic energy spectra.}
\label{compa_evo_eb_ot} \end{figure}

We finally investigate the spectral behavior of our model on
this particular Orszag-Tang flow. 
We respectively plot in  Fig.~\ref{compa_evo_ev_ot} and
Fig.~\ref{compa_evo_eb_ot} the kinetic and magnetic spectra
of both DNS and LES MHD II, at times $t=1$, $t=3$, $t=5$, and $t=10$.

One can observe strong similarities of the modeled spectra with
the directly simulated ones, although small differences appear
at large scales.
 
\begin{figure}[h!]
\includegraphics[width=7cm, height=50mm]{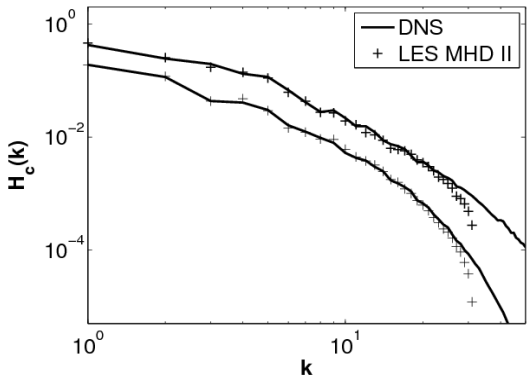}
\caption{Correlation spectra at $t=3$ (top) and $t=10$ (bottom),
for data  {\bf VII} ($256^3$ DNS straight line), and {\bf VIII} 
($64^3$ LES MHD II +).}
\label{compa_evo_ec_ot} \end{figure}

In order to evaluate the effect of the 
model on the cross-correlation, scale by scale, we represented
in Fig.~\ref{compa_evo_ec_ot} its spectra at times $t=3$ and $t=10$
(we show only two times for readability purpose).

We can observe that at large scales which are the most energetic, 
the model reproduces correctly the spectra obtained with the DNS 
at both times. But, close to the cut-off, the model strongly under-estimates
the cross-correlation. This phenomenon, as stated before, is linked to the eddy viscosity and eddy diffusivity
which dissipate the kinetic and magnetic resolved scales, as well as
the cross-correlation at these scales.       
The reconstruction procedure allows to reinject energy
and helicity (when taken into account) at these scales, but not the correlation.
\section{Conclusion} \label{conclu}

In this paper we accomplish two complementary tasks.
We first develop a LES for MHD using the EDQNM equations and
transport coefficients derived in \cite{PFL} but in the non-helical case. 
We then show that not all relevant time scales appearing in the cumulant expansions of the primitive MHD equations are
taken into account in the phenomenological formulation of  \cite{PFL}. Indeed, one can derive 
several new eddy-damping times for
the EDQNM equations, and document how, by using them, one can considerably improve 
the treatment of the magnetic and kinetic energy transfers in the Large Eddy Simulation approach taken in this paper.

A possible extension of this work is to be able to incorporate the effect
of either cross-helicity \cite{grappin82} and of kinetic
and magnetic helicity \cite{PFL} in the evaluation of eddy viscosities
and eddy noise. The fact that the modeling algorithm does not depend on a specified 
inertial index may also be of some help in the case of a high velocity-magnetic field 
correlation when different spectra emerge at high values of the (normalized) $H^C$ cross-helicity \cite{slow}.

Furthermore, with such a model many astrophysical and geophysical flows can be 
studied and perhaps more importantly a vast range of parameters, in particular the magnetic Prandtl number, 
can be examined. Among such problems, the generation of magnetic fields at either low or high 
magnetic Prandtl number is of prime importance, in particular in the former case in view of a 
set of laboratory experiments studying this effect \cite{monchaux}.

\begin{acknowledgments}
This work is supported by INSU/PNST and PCMI Programs and CNRS/GdR Dynamo.
Computation time was provided by IDRIS (CNRS) Grant No. 070597, 
and SIGAMM mesocenter (OCA/University Nice-Sophia).
\end{acknowledgments}

\appendix \label{appen}
\section{EDQNM closure} \label{appenA}
For completeness, we recall here the expression of the EDQNM closure 
equations for the magnetic and kinetic energy without helicity. The first non-helical EDQNM equations were first derived in \cite{KN} but we follow here the notation of \cite{PFL} which gives the free helical closure: 
\begin{eqnarray}
(\partial_t + 2\nu k^2 )E^V(k,t) & =  & \widehat{T}^V(k,t)\\
(\partial_t + 2\eta k^2 )E^M(k,t) & =  & \widehat{T}^M(k,t)
\end{eqnarray}
where the nonlinear transfer terms for the kinetic and magnetic energy,
respectively $\widehat{T}^V(k,t)$ and $\widehat{T}^M(k,t)$ are expressed as:
\begin{eqnarray}
\widehat T^{V}(k,t)& = &
\iint_{{\Delta}_k}\theta_{_{kpq}}(t)S^V(k,p,q,t)dpdq  \ ,\label{tvedqnm}\\
\widehat T^{M}(k,t)& = & 
\iint_{{\Delta}_k}\theta_{_{kpq}}(t)S^M(k,p,q,t)dpdq  \ . \label{tmedqnm}
\end{eqnarray}
Here $\Delta_k$ is the integration domain with $p$ and $q$ such that 
($k,p,q$) form a triangle, and $\theta_{_{kpq}}(t)$  namely the 
triad-relaxation time is expressed as:
\begin{equation}
\theta_{kpq}(t) = \frac{1-e^{-\mu_{kpq}t}}{\mu_{kpq}} \ \ ,
\label{pfl_rate} \end{equation}
with $\mu_{kpq}=\mu_k +\mu_p +\mu_q$ where the $\mu_k$'s are called 
eddy-damping rates and read:
\begin{eqnarray}
\mu_k &=&  + \lambda \left(\int_0^k \!\! q^2( E_q^V+E_q^M) dq
\right)^{\frac{1}{2}} \nonumber \\
& + & \sqrt{\frac{2}{3}}k \left(\int_0^k\!\! E_q^M dq 
\right)^{\frac{1}{2}}+(\nu +\eta )k^2.
\label{mu_PFL} \end{eqnarray}
The constant $\lambda$ can be expressed as a function of the Kolmogorov
constant $C_k$ appearing in front of the kinetic energy spectrum such that:
\begin{equation}
\lambda = 0.218 C_k^{\frac{3}{2}} \ \ ,
\label{lambda} \end{equation}
following \cite{chollet_lesieur}.

The expressions of $S^V(k,p,q,t)$ and $S^M(k,p,q,t)$ can be further explicited 
(with the time dependency of magnetic and kinetic energy spectra 
omitted here) as :
\begin{eqnarray}
S^V(k,p,q,t) & = &\frac{k}{pq}b_{kpq}\left[k^2E^V(q)E^V(p)
-p^2E^V(q)E^V(k)\right]\nonumber \\
& + & \frac{k}{pq}c_{kpq}\left[k^2E^M(q)E^M(p)
-p^2E^M(q)E^V(k) \right] \nonumber \\
& = & S_1^V(k,p,q,t) + S_2^V(k,p,q,t)\nonumber \\
& + & S_3^V(k,p,q,t) + S_4^V(k,p,q,t)  \ .
\label{S^V}
\end{eqnarray}
\begin{eqnarray}
S^M(k,p,q,t) & = &\frac{k}{pq}h_{kpq}\left[k^2E^M(p)E^V(q)
-p^2E^V(q)E^M(k)\right]\nonumber \\
& + & \frac{k^3}{pq}c_{kpq}\left[\frac{k^2}{p^2}E^V(p)E^M(q)
-E^M(q)E^M(k) \right] \nonumber \\
& = & S_1^M(k,p,q,t) + S_2^M(k,p,q,t)\nonumber \\
& + & S_3^M(k,p,q,t) + S_4^M(k,p,q,t)  \ .
\label{S^M}
\end{eqnarray}
In Eqs.~(\ref{S^V}) and (\ref{S^M}) the geometric coefficients $b_{kpq}$,
$c_{kpq}$, and $h_{kpq}$ are defined as:
\begin{eqnarray}
b_{kpq} &=& pk^{-1}(xy+z^3), \quad  c_{kpq} = pk^{-1}z(1-y^2) \ ,\nonumber\\
h_{kpq} &=& z(1-y^2) \ ,\nonumber
\end{eqnarray}
where $x$, $y$, $z$ are the cosine of the interior angles opposite to ${\bf k},{\bf p},{\bf q}$. This completes the description of the EDQNM closure for MHD as developed in \cite{KN,PFL}. 
The helical case, dealt with in \cite{LESPH} for a pure fluid and in \cite{PFL}
from the EDQNM standpoint, will be studied in a forthcoming paper when coupling to a magnetic field is involved.

\section{A more general eddy damping}  \label{appenB}
In \cite{PFL}, the eddy damping term is built on a phenomenological ground; namely, one argued about the necessity of introducing the Alfv\'en time scale in the damping coefficient, without actually referring to the set of cumulant expansion equations. This change alone, from a traditional hydrodynamic EDQNM closure, led to energy spectra that differ from the Kolmogorov case, with a $k^{-3/2}$ law in the uncorrelated case, and with a $E^{\pm}(k)\sim k^{-m^{\pm}}$ in the correlated case \cite{grappin83}, with $m^++m^-=3$; here, $E^{\pm}(k)$ are the energy spectra of the Els\"asser variables 
${\bf z}^{\pm}={\bf v}\pm {\bf b}$.

However, when examining the succession of equations for the higher-order moments, and keeping the total correlation between the velocity and the magnetic field equal to zero to simplify the algebra, a more complex structure emerges, which may help the modeling of the MHD dynamics to be closer to the DNS than the results shown in Fig.\ref{compa_evo_pfl}. There are in fact four groups of terms in the closure equations.
The first group corresponds to the pure fluid case and can be written symbolically as:
\begin{equation}
\left(\frac{\partial}{\partial t} + \nu (k^2 + p^2 +q^2)\right )<uuu>
\simeq  k <uuuu> \ .
 \label{Tg1}\end{equation}
 This leads, as usual, to two characteristic times written here as:
$\tau_D^{VV}=\left(\nu(k^2 + p^2 +q^2)\right)^{-1}$
and $\tau_{NL} =\left(ku \right)^{-1}$.
The second group writes symbolically again as:
\begin{equation}
\left(\frac{\partial}{\partial t} + \nu k^2 + \eta (p^2 +q^2)\right )<uuu>
 \label{Tg2}\end{equation}
$$\simeq  k <uuuu> + k <bbuu>\ .$$
Here, two new times can be extracted, namely a dissipative time
$\tau_D^{VM}=\left(\nu k^2 + \eta (p^2 +q^2)\right )^{-1}$\ ,
 and
$\widetilde{\tau}_A  = u(kbb)^{-1}$, a modified Alfv\'en time.

The third group of closure terms is of the following type:
\begin{equation}
\left(\frac{\partial}{\partial t} + (\eta k^2 + \eta p^2 +\nu q^2)\right )<bbu>
 \label{Tg3} \end{equation}
 $$ \simeq  k <bbbb> + k <bbuu> \ .$$
\noindent and finally the fourth group:
\begin{equation}
\left(\frac{\partial}{\partial t} + (\eta k^2 + \nu p^2 +\eta q^2)\right )<bbu>
 \label{Tg4} \end{equation}
 $$ \simeq  k <bbbb> + k <bbuu> \ .$$
Again, the following new characteristic dissipative times can be a priori deduced from these two groups (with similar nomenclatures as before):
$\tau_D^{MM}=\left(\eta k^2 + \eta p^2+\nu q^2 \right)^{-1}$, and
$\tau_D^{MV}  =  \left(\eta k^2 + \nu p^2+\eta q^2\right)^{-1}$.

In conclusion, a careful examination of the cumulant equations has led to the adjunction of several new times, distinguishing between magnetic and kinetic energy transfer 
as well as the different quantities entering the transfer terms.

Note that the new modified Alfv\'en time is finally expressed as:
\begin{equation}
\widetilde{\tau}_A^{-1}(k) =  C_A \left(\frac{\int_0^k E^M(q)dq}
{\int_0^k E^V(q)dq}\right)^{\frac{1}{2}} 
\left[\int_0^k q^2 E^M(q)dq 
\right]^{\frac{1}{2}} \ .\end{equation}
It incorporates the lack of equipartition between the kinetic and magnetic energy that is often observed, and this for example should also alter the dynamics, in the early (kinematic) phase of the dynamo problem. The non-linear time has the classical expression built only on the velocity field:
\begin{equation}
\tau_{NL}^{-1}(k) =  \lambda \left[\int_0^k q^2 E^V(q)dq \right]^{\frac{1}{2}} \ .
\end{equation}
Finally, we numerically estimated the value of the Alfv\'en time constant
$C_A=0.8$. This point will need further study as we extend the number of flows that are tested with this LES.
The constant $\lambda$ is determined through the relation
[\ref{lambda}] in Appendix A. The model has thus two open parameters that can be evaluated once the constants appearing in front of the energy spectra in MHD are determined.

Also note that the way the dissipation coefficients are taken into account may well affect the results when the magnetic Prandtl number differs substantially from unity
unless possibly when both the kinetic and magnetic Reynolds numbers are very large because of the effect of renormalisation of transport coefficients \cite{fournier}. 

Similarly to equation (\ref{pfl_rate}) for the eddy damping rate in \cite{PFL}, we define generalized rates as:
\begin{equation}
\theta_{kpq}^{XY}(t) = \frac{1-e^{-\mu^{XY}_{kpq}t}}{\mu^{XY}_{kpq}} \ \ ,
\label{new_rate}\end{equation}
with $\mu^{XY}_{kpq}=\mu^{XY}_k +\mu^{XY}_p +\mu^{XY}_q$ and with $XY$ standing for either $VV$, $VM$, $MV$ or $MM$ and with:
\begin{equation}
\mu_k^{VV} = \left(\tau_D^{VV}\left(k\right)\right)^{-1} 
+ \left(\tau_{NL}\left(k\right)\right)^{-1} \ ,
\label{muvv}
\end{equation}
\begin{equation}
\mu_k^{VM} = \left(\tau_D^{VM}\left(k\right)\right)^{-1} 
+ \left(\tau_{NL}\left(k\right)\right)^{-1} 
+ \left(\widetilde{\tau}_A\left(k\right)\right)^{-1} \ ,
\label{muvm} 
\end{equation}
\begin{equation}
\mu_k^{MV} = \left(\tau_D^{MV}\left(k\right)\right)^{-1} 
+ \left(\tau_{NL}\left(k\right)\right)^{-1} 
+ \left(\widetilde{\tau}_A\left(k\right)\right)^{-1} \ ,
\label{mumv} 
\end{equation}
and
\begin{equation}
\mu_k^{MM} = \left(\tau_D^{MM}\left(k\right)\right)^{-1} 
+ \left(\tau_{NL}\left(k\right)\right)^{-1} 
+ \left(\widetilde{\tau}_A\left(k\right)\right)^{-1} \ .
\label{muvm}\end{equation}

\end{document}